*Synthesis of novel Mesoporous Fullerene and its application to electro-oxidation of methanol.*

Sujit Kumar Mondal[a, b†]

[a] Past Affiliation-Nano-ionics Materials Group, Fuel Cell Materials Center, National Institute for Materials Science, 1-1 Namiki, Tsukuba Ibaraki, 305-0044, Japan

[b†] Present Affiliation-Biomaterial And Health Science Centre, University of Texas, Houston, Texas-77030,USA. Email:- Sujit.K.Mondal@uth.tmc.edu

**Abstract**

One pot synthetic technique was used to synthesize mesoporous fullerene using highly ordered mesoporous template such as KIT6-150 and SBA-15-150. Highly ordered porous fullerene with cross linkage between the fullerene molecules was obtained at high temperature 900 $^0$C. This type of highly ordered and high surface area mesoporous fullerence was used as catalyst to study anodic performance in sulphuric acid medium. Finally, it was used to test its catalytic activity for electrochemical methanol oxidation. For anodic performance and methanol oxidation experiment, mesofullerene was impregnated with pt-nano particles. Mesoporous Fullerene-Pt nano-composite showed quite an unusual electrochemical property for electro oxidation of methanol and good anodic performance in acidic electrolyte. The anodic performance was much better than the commercially available carbon black. This short communiation is a report mentioning the synthetic strategy to prepare mesofullerene and electrode preparation to study its electrochemical performance. This mesofullerene is potential catalyst for fuel cell application.

## Introduction

Ordered mesoporous materials (OMMs) are of particular interest in many fields because of their favorable structural properties, which meet many application requirements. One of the aspects is that their pore size can be controlled [1]. Porous material (carbon) with nanoscale pore size prepared from periodic inorganic silica templates have been receiving much attention because of their versatility and shape selectivity. These properties of mesoporous material will be helpful for nano-reactor, battery electrodes, capacitor, energy storage device and fuel cell electrodes. Nano-porous well ordered materials are very important for the application for catalysis, adsorption and fuel cell due to their excellent textural characteristics [2,3]. Template technique to prepare the nano-porous material (carbons) with well ordered structured is an unique and well known methodology [4-12]. Appropriate source for the target material was first impregnated inside the template structure of ordered mesoporous channel followed by solidification and removal of template will give rise to nanopoprous material [4-13]. Synthesis of nanoporous carbon using various mesoporous templates and verification of their chemical/electrochemical activity was reported in literature [1,14].

Mesoporous material having various applications when their pore sizes can be tuned with their textural property. Multiscale pores controlled (with larger or smaller sizes) and well connected pores would be beneficial for specific application for example, catalysis and fuel cell [14]. Another most rapid developing direction of contemporary chemical physics is associated with the discovery of fullerenes, which represents a novel allotrope of carbon [15-20]. Carbon atoms in fullerene molecules are located at the vertices of regular hexagons and pentagons, which cover in a regular manner the surface of a sphere or spheroid. The most spread and elaborated molecules belonging to the fullerene family is $C_{60}$, whose structure is a regular truncated icosahedrons. The surface of this molecule is constructed from twenty regular hexagons and twelve regular pentagons so that each pentagon is adjacent only to hexagons, whereas each hexagon is adjacent to three pentagons and three hexagons alternately. The fullerene family also includes along with $C_{60}$ the molecules $C_{70}$, $C_{76}$, $C_{78}$, $C_{84}$, etc., distinguished by a lower symmetry and larger number of hexagons on the surface. Thus, the fullerenes form a unique class of molecules having a closed two-dimensional structure. Here author has showed an experimental strategy to synthesize in-situ novel mesoporous fullerene using highly ordered mesoporous KIT-6 and SBA-15 templates. The newly synthesized mesoporous fullerene having highly ordered texture, high surface area 310.5m$^2$/g with nano pores of 3.5 nm diameters (obtained from surface area measurement). Another meso fullerene obtained from SBA-15 having lower surface area (100m$^2$/g) with nanopore of 6nm diameter, this meso-fullerene showed an excellent electrochemical performance with unique feature of randomly distributed porous bundle of rod shaped micro-structure. The performance of this meso fullerene was examined, its electrochemical performance dominated over the performance of 20wt% pt loaded carbon black, reported in literature [14]. Unique features of mesoporous fullerene are having tremendous applications for PEMFc such as DMFc. A high oxidation current was observed for 10 wt% and 20wt% Pt loaded with mesoporous fullerene during anodic performance and electrochemical oxidation of methanol.

**Experimental**

Here the design of synthesis of mesoporous fullerene was described; target material (fullerene) was impregnated with proper stoichiometric quantities inside mesoporous template KIT6-150 and SBA-15-150. Fullerene was first dissolved in 1,2,4 trinitro benzene in the appropriate proportion and mixed thoroughly with the KIT6-150 then the sample was transferred in the carbonization chamber under 900 $^0$C in presence of nitrogen atmosphere for 6 hours. The ordered mesoporous material (KIT6-150 template) was impregnated by precursor fullerene, followed by intertwined of fullerene inside the mesoporous channel was obtained at 900 $^0$C in presence of nitrogen atmosphere. After 6hrs of heat treatment, it was cooled down to room temperature and the mesoporous fullerene/template composite was washed thoroughly by 5 wt% of HF to remove silica. Then the sample was dried properly at 120 $^0$C and ready for characterization. Same procedure was followed for another template, namely SBA15-150, mesoporous fullerene with rod-shaped morphology was obtained using SBA15-150 template.

The dried sample was used first for XRD measurement using multisampler rigaku diffractometer from 0 to 80 degree for 2θ value. Powder sample was used for Quantochrome Autosorb 1 sorption analyzer to measure the $N_2$ adsorption and desorption isotherm. Surface area and pore diameter were calculated by BET and BJH method using the instrument software. The mesoporous fullerne sample was used for SEM and TEM to find out morphological information. Machine named FESEM Hitachi-S 4800 and TEM JEOL JEM-2000EX2 were used for SEM and TEM experiments. For anodic performance and electro oxidation of methanol, 10wt%, 20 wt% of platinum loaded mesoporous fullerene were used as active electrode material. For electrochemical measurement mesoporous fullerene with two different morphology (obtained from two different templates) were dispersed in the ethanol solution, the mixture was stirred for 1hr to complete the mixing of platinum salt and mesoporous fullerene. Once again it was stirred for 6hours in presence of $N_2$ gas purging at room temperature and lastly the dried sample was heated at 400 $^0$C to reduce the platinum in a stream of gas containing $H_2$ (10%) and $N_2$ (90%) for 2 hours. The composite pt-meso-fullerne was dispersed in methanol (2mg/ml) and 5 µL solution was dispersed over gold electrode having area 0.2cm$^2$. The catalyst coated gold electrode was used for anodic performance and methanol oxidation study. The electro oxidation of methanol and anodic performance were studied using CH760 chemical analyzer. Each cyclic voltammogram was recorded at 50mV s$^{-1}$ scan rate. Every electrochemical data was collected after the completion of 50cycles of cyclic voltammetry experiment, for each new experiment new electrode was used; all experiment was carried out in $H_2SO_4$ medium.

## Results and Discussion

XRD measurement was performed with dried samples after removal of silica template, it displayed a unique XRD pattern. This XRD pattern was described in Figure1A and 1B in supporting section of the manuscript. Peak at low angle 2θ ~ 1.2 and another small peak around 2θ ~ 2 describing the presence of ordered mesoporous fullerene like ordered mesoporous carbons [14]. The other peaks at higher angle also resembled with peak for $C_{60}$ fullerene mainly 2θ ~ 15 and 2θ ~ 30 was described in figure 1B. The broad peaks at 2θ ~ 26 and 2θ ~ 44 possible due to the overlap of peaks indicating a cross linking of fullerenes was occurred at 900 $^0$C.

The mesoporous fullerene is having a very well ordered microstructure, SEM pictures described the microstructure of mesofullerene. Microstructure of mesofullerene, obtained from two different silica templates were very different and unique. It was found that mesoporous fullerene having porous morphology with pore varying from 200 nm to 400 nm, described in figure.1(A) and 1(B). This type of unique microstructured mesoporous fullerene exhibited globular shape particles with less than 100nm size, possible vertical orientation with respect to a plane of paper. This type of microstructure was obtained as a results of using highly ordered 3 dimensional interconnected KIT6 template. The mesoporous fullerene with rod-shaped morphology was obtained from 2-D ordered SBA15 template, was described in fig.1(C), 1(D). SEM pictures it was clear that the dimension of the rod was unevenly distributed. The smaller rod are having length of 8-10 micron and diameter around 1.5 micron, long rod shaped mesoporous fullerene having length of 25-28 micron and dia around 2 micron. These bundle of rods created highly porous microstructured, these type of highly porous microstructure can produce facile capillary rise action path for pt diffusion. This is a valid reason for mesoporous fullerene to show a very promising result for electrochemical analysis.

Detailed transmission electron microscope investigation gave further insight into the mesoporous structure. TEM bright field images were recorded at low and high magnifications and were described in figure 2(A), 2(B), 2(C) and 2(D) in the supporting section. The size of the mesoporous fullerene-pt composite aggregate varied from 20 nm to 50 nm. It was clear from the figure 2(A) that the fullerene aggregate consisted of ultra fine particles of Pt confirming the efficient loading of Pt particles inside mesofullerene structure; this mesofullrene was obtained using the KIT-6 porous template. The fringes were clearly visible in figure 2(B), the average particle size of Pt measured from TEM images (Fig 2C) was 6 nm with nano pore of 3 nm in diameter. Interestingly, ordered Pt nano particles were often observed within the singly fullerene-Pt cluster, which resulted in ordered porous structure of Pt-fullerene nanocomposite. The size of the Pt particles was about to 6nm in the case of mesofullerene obtained from SBA-15 templete, figure 2(D). The mesoporous fullerene not only having nano dimension but also having a very high surface area compare to original fullerene molecule 0.9 m$^2$/g [9]. The high surface area was confirmed by the nitrogen adsorption and desorption measurement, reported in supporting section figure 4(A) and 4(B). Mesoporous fullerene was obtained from KIT-6 having surface area of 310.5 m$^2$/g with pore diameter 3.53 nm, mesofullerne was obtained from SBA-15-150 template having surface area of 100 m$^2$/g with large pore diameter (6nm). This very high surface area can build up a very high double layer charging current during; high double layer charging current was observed during electrochemical anodic performance study. This double layer charging current was observed for 10wt% pt loaded mesoporous fullerene and also with 20wt% pt loaded meso-fullerene composites. This high double layer charging current for 10wt% pt loaded mesoporous fullerene (obtained from SBA-15-150 template) was reported in figure 2(A) and 2(C) in present research article. Electrochemical performances of mesoporous fullerene (obtained from KIT-6 template) with 10wt% pt-loading were described in figure 2(B) and 2(D). This effect was also observed for other high wt% of pt loaded mesoporous fullerene-pt composite, during anodic performance study and results were described in figure 3(A) and 3(B) of the supporting section of the present article.

The anodic performance of mesoporous fullerene was carried out in 0.5 M H$_2$SO$_4$ solution. Before electrochemical experiment fullerene was loaded with 10 and 20 wt % of platinum then pt-mesoporous fullerene composite were dissolved in required amount of EtOH followed by a sonication for 30 min. 5μL of the solution mixture was transferred over a gold electrode to study cyclic voltammetry experiment with a sweep rate of 50 mV s$^{-1}$scan rate, potential was varied from -0.2 to 1.0 V with respect to Ag/AgCl electrode. This experiment repeated for 50 cycles to obtain a reproducible result. The cyclic voltammetry data of anodic performance study showed a very promising current value for the both 10 wt% and 20 % wt of platinum loaded mesoporous fullerene [22]. Composite of Pt-mesoporous fullerene composed of high surface area which enhanced ions adsorption followed by diffusion, as a result of which, a high oxidation current was observed during electrochemical methanol oxidation experiment. The conjugation between the fullerene molecules in mesoprous fullerene not only produced very high surface it also produced a very high electronic conductivity due to availability of the π- orbital. High surface area and π-π interaction of mesoporous fullerene were possible cause to show an excellent performance during anodic electrochemical reaction.

In anodic performance study, hydrogen desorption peak was observed at -0.1347V, this peak due to weakly bonded hydrogen at the platinum surface and -0.0646V peak was due to strongly bonded hydrogen at the surface[23]. This phenomenon was described in figure 2(A) and 2(B) in the article section, $H_W$ and $H_S$ were recognized to be the peaks for weakly adsorbed hydrogen and strongly adsorbed hydrogen at the surface. This phenomenon was observed for all the mesofullerene catalysts loaded with various pt wt%, values were tabulated in the table-1. 20wt% pt loaded catalyst was also examined for anodic performance with methanol oxidation study and values were reported in the table-1. All electrochemical figures for 20wt% pt loaded catalyst were described in the supporting section with figure 3(A) and 3(B). In the following table potential values for mesoporous fullerene impregnated with different wt% of pt was tabulated. The hydrogen desorption potential values were tabulated from anodic performance study together with anodic oxidation current values of methanol oxidation experiment. The tolerance value of the catalyst for methanol oxidation was calculated from the ratio of forward oxidation current and backward reduction current values of methanol oxidation experiment.

Single electrode measurements were carried out for all the catalysts to study the methanol oxidation. The fullenere obtained from mesoporous SBA15 template having a peak current value for oxidation 1.616 mA cm$^{-2}$ with reverse anodic peak at 0.625 mA cm$^{-2}$. Forward anodic peak was observed at 0.593V and reverse peak at around 0.38V. Forward anodic peak was due to the oxidation of methanol and backward peak was attributed for removal of incompletely oxidized carbonaceous species formed in the forward scan. Hence, the $I_a/I_b$ can be used to describe the catalyst tolerance limit for the accumulation of carbonaceous species. Low tolerance value means poor oxidation of methanol to carbon dioxide and excessive accumulation of carbonaceous species on the catalyst surface and high tolerance value means the reverse case. The tolerance limit for this catalyst for electrochemical oxidation of methanol was found to be 2.6 and for mesoporus fullerene obtained from KIT6 was 2.8. The pt-mesoporous fullerene showed high tolerance value for methanol oxidation, high tolerance value indicate better methanol oxidation and less accumulation of carbonaceous species on the catalyst surface. Author believed that pt diffusion inside the surface of mesofullerene and availability of electro-active surface area of pt could be a valid reason but quantitative understanding is still under progress. In order to explain the reason behind high oxidation current for 10 wt% meso-fullerne obtained from SAB-15, it was important to consider the structure of the parent template (2-diemnsional hexagonal structure). Parent template was having 2 –dimensional rod shaped hexagonal structure which untimely produced beautiful rod shaped microstructure for mesofullerene and led to a facile capillary rise phenomenon for pt diffusion, this possibly created beautiful ultrafine pt- nano-particle with single pt-fullerne nano-cluster. This phenomenon was clearly observed from the TEM images of mesoporous fullerne obtained from KIT-6 template. This unique porous feature definitely produced more site specific growth of highly ordered pt nano-particles inside the mesoporous rod-shaped fullerene structures obtained from KIT-6, which in return showed a high current value for electrochemical oxidation of methanol. Author believed that the presence of ultrafine pt-nanoparticles inside the mesopores of mesoporous fullerene can effectively remove the carbonaceous species formed during methanol oxidation process; this would open up effective surface area of the catalyst for overall electrooxidation of methanol. Ultrafine pt-nano particles was clearly observed from TEM images of mesoporous fullerene. This unique porous feature definitely produced more site specific growth of highly ordered pt nano-particles inside rod-shaped and globular shaped mesoporous fullerene.

In addition, 20wt % platinum loaded mesoporous fullerene was also used to study the methanol oxidation reaction in presence of 0.5 (M) $H_2SO_4$ medium. A high current value of 0.58 mA at potential of 0.6 V with respect to Ag/AgCl electrode was described in figure 6(B). The tolerance limit for the catalyst was calculated from the ratio of $I_a$ and $I_b$, a small tolerance value of 1.26 was obtained from the present anodic and cathodic current peak. The enhanced oxidation current was due to enhancement of pt content inside mesopsorous-fullerene but high % of pt definitely produced agglomeration inside the mesopores fullerene structure, which was responsible for a low tolerance limit for the catalyst. The first hand results of Pt-mesofullerene composite were promising enough to explore the mesoporous fullerene as a potential catalyst for DMFC [22].

**Conclusion:-** The mesoporus fullerene was successfully synthesized from well ordered mesoporous KIT-6 150 and SBA15. The aim of the synthesis was to achieve a well ordered, high surface area π- conjugated material and the goal was successfully entangled. From the TEM images it was very clear that pt super lattice was obtained with well ordered array of mesoporous fullerene. The novel material showed a very high surface area compare to normal fullerene molecule and high conductivity due to π- π interaction was resulted from cross linkage. Electrochemical experiments showed a marvelous result for anodic performance study and also direct methanol oxidation process. This was first hand result about mesoporous fullerene obtained in the laboratory scale and all preliminary results were described here as a manuscript.

**Acknowledgement:-** Author acknowledges National Institute for Material Science, Namiki, Japan for his postdoctoral fellowship and funding because this work was carried out in NIMS, japan during his postdoctoral period. Author thanks A. Vinu for giving a space in his synthesis lab and Dr. T Mori to allow his electrochemistry laboratory. Author acknowledges Dr. P Srinivasu for his useful discussion about mesoporous templates synthesis and Dr. Balu for his support during SEM experiment.

| Samples | Anodic peak for hydrogen desorption | Methanol oxidation current value with tolerance ratio |
|---|---|---|
| Meso-fullerene obtained from SBA-15 template. 10wt % of pt | $H_W$ -0.135V, $H_S$ = -0.065V | 1.616 mA cm$^{-2}$ at 0.593 V. $I_a/I_b$ = 2.585 |
| Meso-fullerene obtained from KIT-6 template. 10wt % of pt | $H_W$ -0.132V, $H_S$ = -0.062V | 0.504 mA cm$^{-2}$ at 0.609 V. $I_a/I_b$ = 2.7937 |
| Meso-fullerene obtained from KIT-6 template. 20wt % of pt | $H_W$ -0.133V, $H_S$ = -0.058V | 3.0 mA cm$^{-2}$ at 0.595V. $I_a/I_b$ = 1.26 |

Table-1 Potential values for hydrogen desorption during anodic performance study, current density for methanol oxidation and the tolerance limit of catalysts for methanol oxidation reaction.

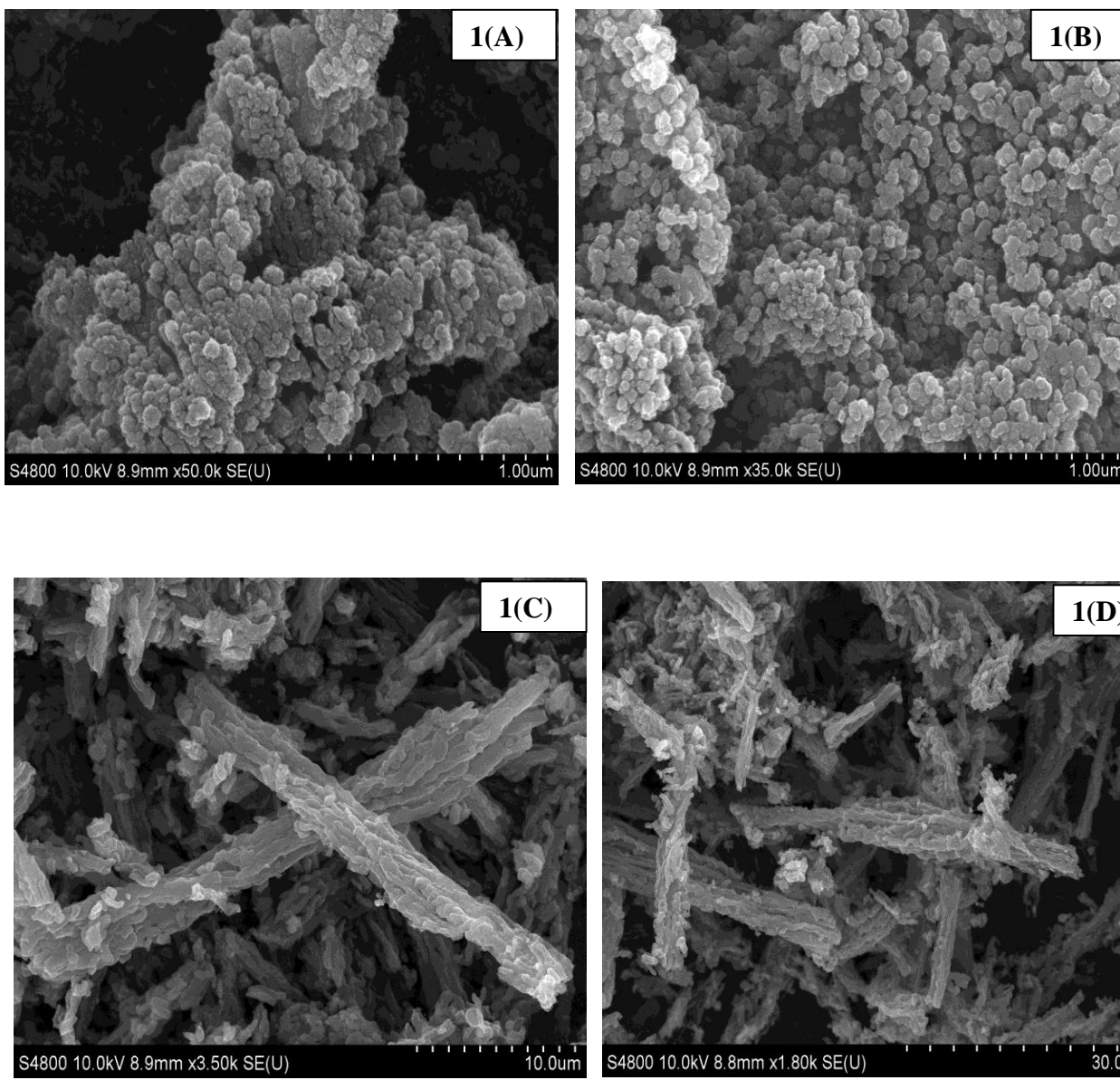

**Fig. 1(A)**, **(1B)** SEM Picture of mesoporous fullerene obtained from mesoporous KIT6 template.

**Fig.1(C), 1(D)** SEM picture of mesoporous fullerene obtained from 2-D mesoporous template of SBA-15.

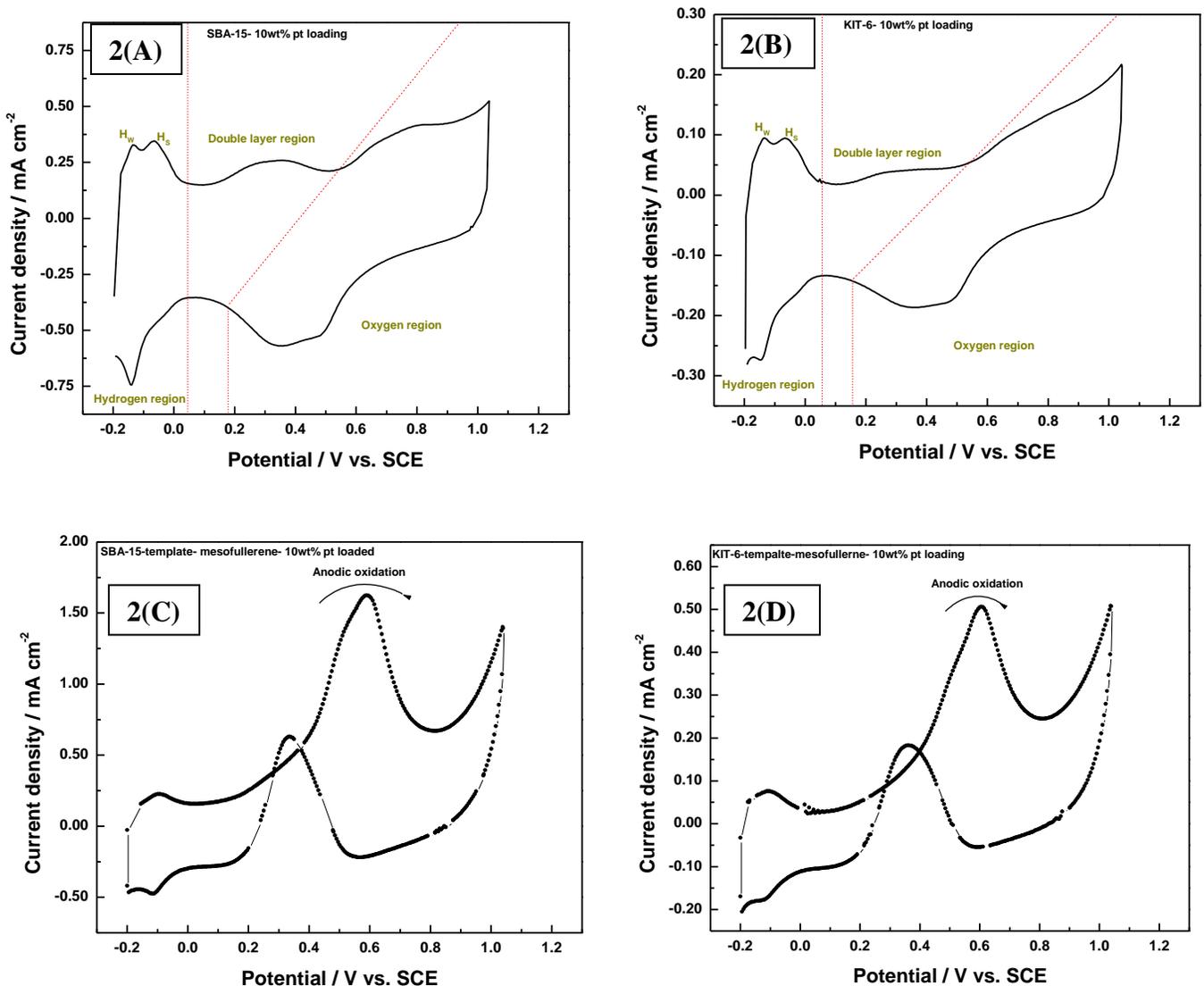

**Fig. 2 (A), (B)** Anodic performance of mesoporous fullerene obtained from KIT6 template & SBA-15 template. The anodic performance of meso fullerne was carried out with 10wt% loading of pt. **Fig.2 (C), (D)** Electro oxidation of methanol by 10wt% pt loaded mesoporous fullerene. Figure **(C)** Mesoporous fuellerene obtained from SBA–15 template and figure **(D)** mesoporousfullerene obtained from KIT-6 template

**Supporting Documents:-**

**Experimental**

Mesoporous fullerene was prepared by the solid-state reaction using proper stoichiometric quantities of mesoporous template KIT6-150 and fullerene. Fullerene was first dissolved in 1,2,4 trinitro benzene in the appropriate proportion and has been mixed thoroughly with the KIT6-150 then the sample has been transferred in the carbonization chamber under 900 $^0$C in presence of nitrogen atmosphere for 6 hours. Here the ordered mesoporous material (KIT6-150 template) was impregnated by precursor fullerene, followed by cross linking of fullerene at 900 $^0$C in presence of nitrogen atmosphere. After the mesoporous fullerene was formed sampled was washed thoroughly by 5 wt% of HF to remove silica. Then the sample was dried properly at 120 $^0$C and ready for characterization. Same procedure was followed for SBA15 template to obtained mesoporous fullerene with nano-rod microstructure.

The dried sample was used first for XRD measurement using multisampler rigaku diffractrometer from 0 to 80 degree 2θ value. Next the powder sample was transferred into Quantochrome Autosorb 1 sorption analyzer to measure the $N_2$ adsorption and desorption isotherm. Surface area and pore diameter were calculated by BET and BJH method using the instrument software. The mesoporous fullerne sample was used for SEM and TEM to find out morphological information. For SEM experiment, machine named Hitachi-S 4800 FESEM was used and for TEM experiment TEM JEOL JEM-2000EX2 was used. The anodic performance and electro oxidation of methanol both 10 and 20 wt% of platinum loaded mesoporous fullerene were studied, fullerene used for electrochemical study were obtained from KIT-6 and SAB-15 template. Electrochemical measurements were carried out for two different meso-porous fullerene obtained from two different silica template. Meso-fullerene was dispersed in the ethanol solution, followed by stirring. Required amount of platinum salt was added to solution. The mixture was stirred for 1hr until the complete mixing of platinum salt and mesoporous fullerene was achieved. Next, it was stirred for 6hours in presence of $N_2$ gas purging at room temperature to achieve the black color powder residue. Finally, the balck colour powder was heated up at 400 $^0$ C to reduce to platinum in presence of a stream of gas containing $H_2$ (10%) and $N_2$ (90%), for 2 hours. The composite of pt-meso-fullerne was dispersed in methanol (2mg/ml) and 5 μL of that dispersed solution was added over the gold electrode, with an active geometric area of $0.2cm^2$. Anodic performance and electro oxidation of methanol were studied in acidic medium of 0.5 (M) $H_2SO_4$ using CH760 chemical analyzer instrument.

**Figures with captions: -**

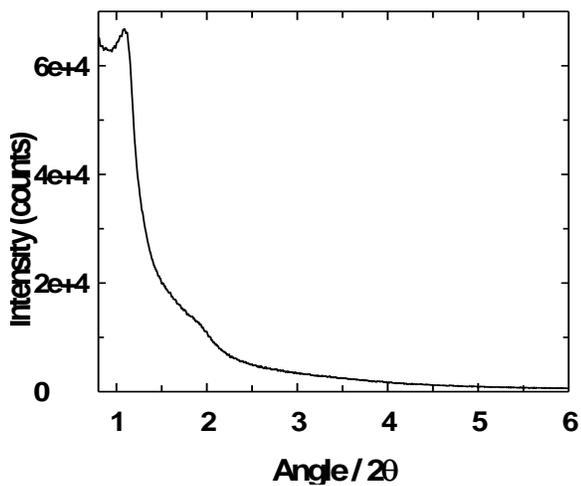
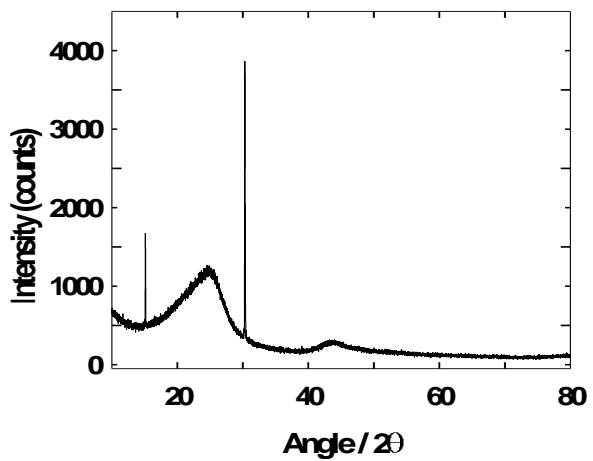

**Fig. 1A**  **Fig. 1B**

Fig.1A and Fig.1B XRD experiment was carried for mesoporous fullerene obtained from KIT6 template, in Rigaku Multisampler Diffractometer machine operated at 40 kV and 40 mA, CuKα = 1.54 A$^0$. Fig.1A explained the pattern related to small scale range of 2θ (0-6) and Fig.1B explained the pattern related to larger scale range of 2θ (10-80).

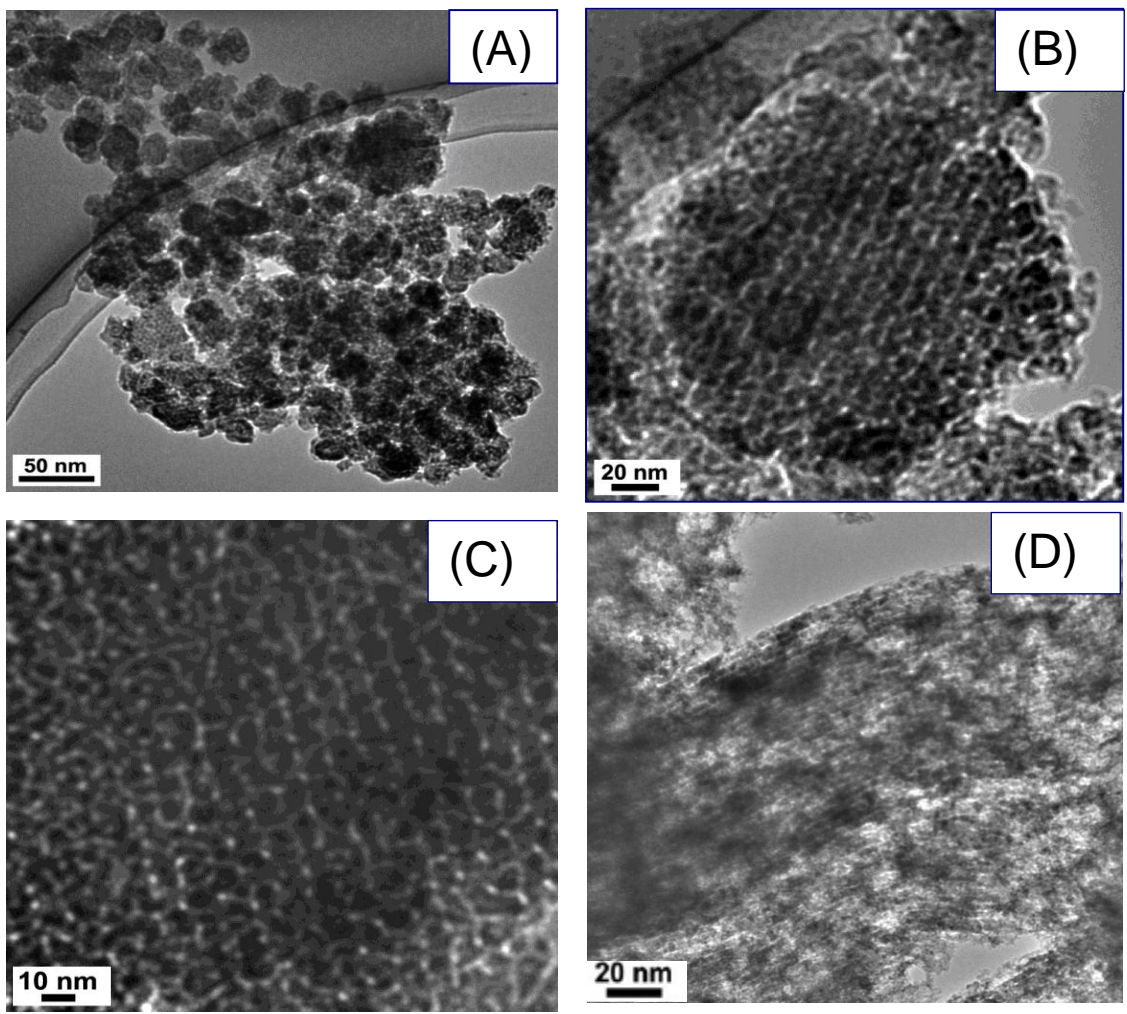

**Fig.2** TEM images of mesoporous fullerene-pt (10wt%) nano composites. HRTEM images were obtained from TEM JEOL JEM 2000EX2 using a copper grid as sample holder. Fullerene obtained from KIT-6 template was described in figure 2(A), 2(B), 2(C) and from SBA-15 template in figure 2(D).

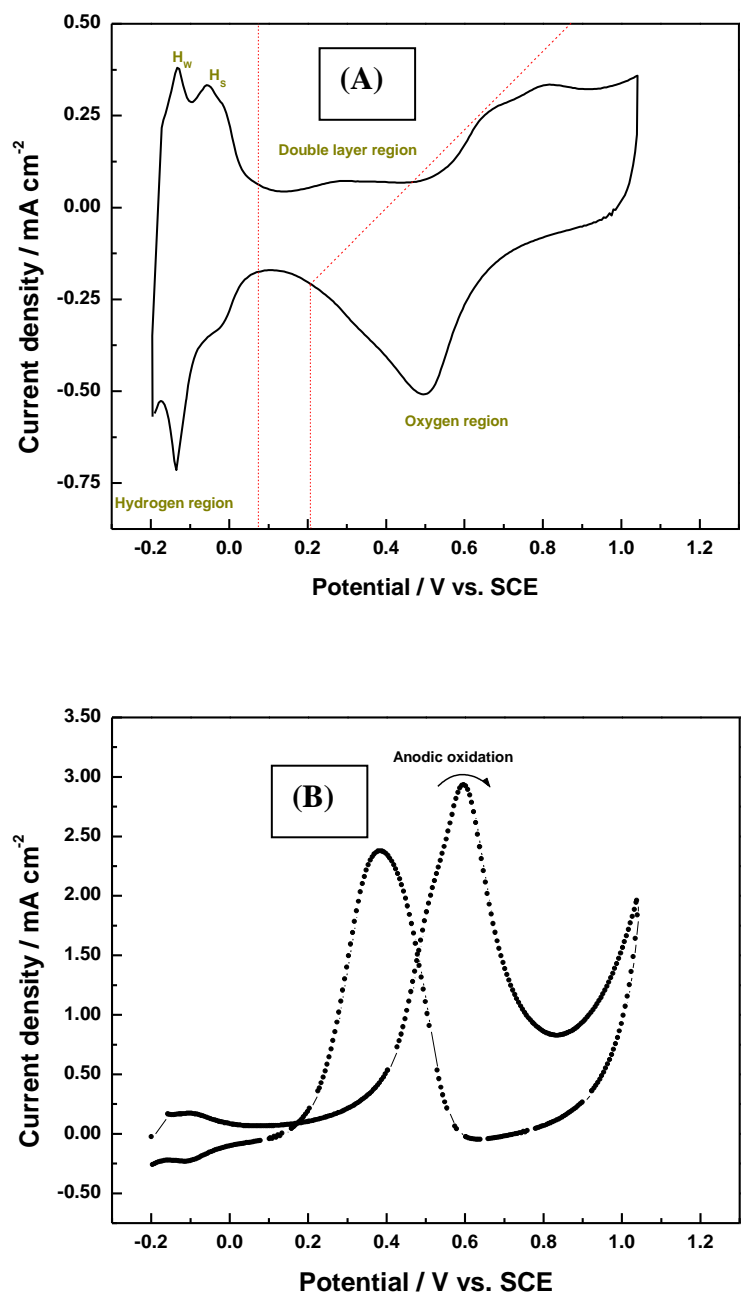

**Fig. 3 (A)** Anodic performance of mesoporous fullerene obtained from KIT6 template. The anodic performance was carried out with 20wt% loading of pt with meso fullerne. **Fig.3 (B)** Electro oxidation of methanol by 20wt% pt loaded meso fullerene.

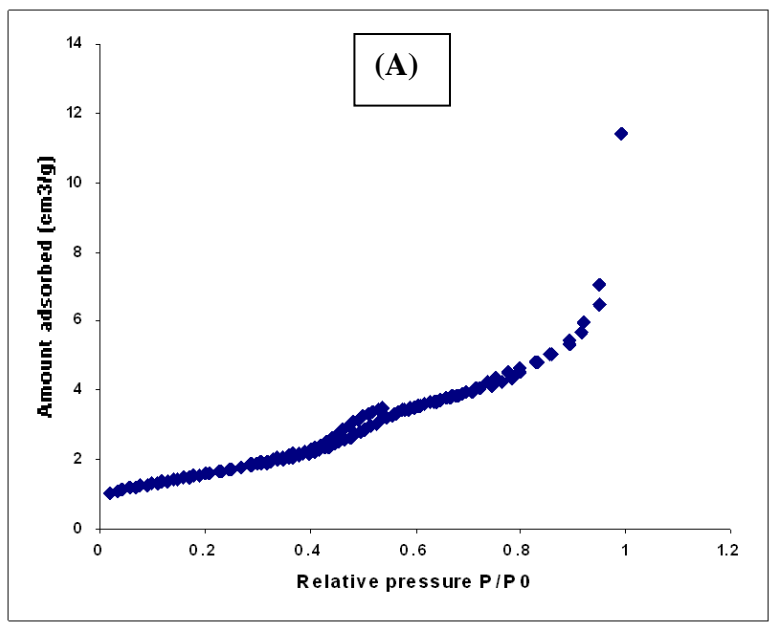

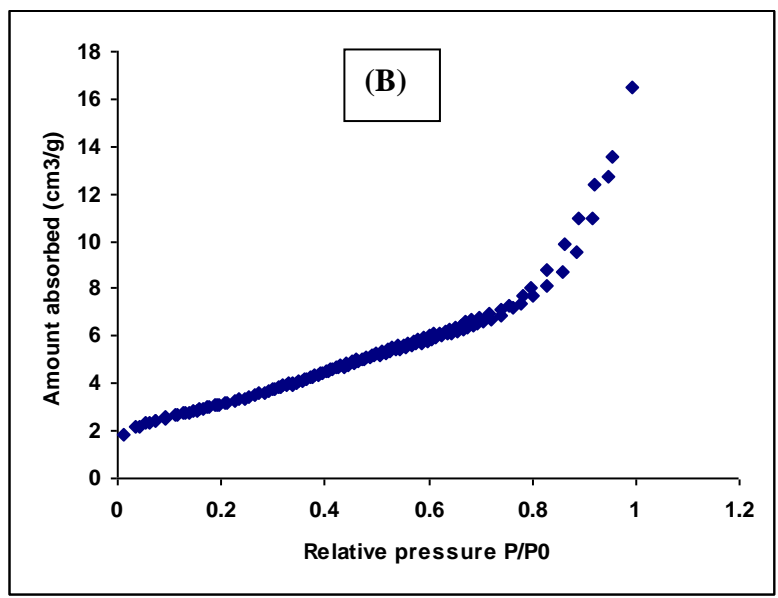

**Fig. 4 (A) and (B),** two different BET isotherm were measured for meso porous fullerene, obtained from two different template namely KIT6 and SBA15.